# Passive radiative temperature regulator: principles and absorption-emission manipulation

HAIPENG ZHAO,[1,2] YAOHUI ZHAN,[1,2,5] SHULIANG DOU,[3] LIANG WANG,[4]

YAO LI,[3] AND XIAOFENG LI[1,2,6]

[1]*School of Optoelectronic Science and Engineering & Collaborative Innovation Center of Suzhou Nano Science and Technology, Soochow University, Suzhou 215006, China.*
[2]*Key Lab of Advanced Optical Manufacturing Technologies of Jiangsu Province & Key Lab of Modern Optical Technologies of Education Ministry of China, Soochow University, Suzhou 215006, China.*
[3]*National Key Laboratory of Science and Technology on Advanced Composites, Harbin Institute of Technology, Harbin 150001, China.*
[4]*State Key Laboratory of High Performance Ceramics and Superfine Microstructure, Shanghai Institute of Ceramics, Chinese Academy of Sciences, Shanghai 200050, China.*
[5]*Corresponding author:Y Zhan (yhzhan@suda.edu.cn)*
[6]*Corresponding author:X Li (xfli@suda.edu.cn)*

**Abstract:** As a representative device exploiting both the solar energy and the radiative cooling of deep-sky, the radiative temperature regulator (RTR) could switch between heating and cooling modes self-adaptively at different temperatures. However, the concept of RTR is challenging to be implemented due to the intense parasitic absorption in phase-changing layers. Here, based on the theoretical framework of energy conservation, we quantitatively reveal the intrinsic relationships between solar heating and radiative cooling, especially addressing the fundamental limiting factors, including the parasitic absorption and the spectral emission selectivity, as well as the dynamic responses of the phase-changing device under various operating conditions. The investigation presents more insight into the underlying physics of RTRs and provides feasible architectures for realizing such a kind of new functional device.

## 1. INTRODUCTION

While the sun is a heat source with a constant temperature of ~6000 K, the deep-sky can be deemed a tremendous cold reservoir with a ~3 K [1–3]. Both resources have long been of interest to researchers, especially in recent years, as green energy has become increasingly advocated. For harvesting solar energy, many technologies are exploited to convert solar light directly to electricity, thermal, or chemical energy, such as photovoltaic conversion [1,4], solar thermal collection [5,6], solar seawater desalination [7–9], and solar water splitting [10,11]. On the other hand, the deep-sky is mainly exploited by using radiative cooling, which is a thermodynamic mechanism that can dissipate the surplus thermal energy of the terrestrial object and thus realizes a steady temperature beneath the ambient circumstance even under the direct sunlight illumination [12–15]. For fully exerting the potential of radiative cooling, Granqvist et al. showed multiple dielectric thin-film designs with highly selective thermal emittance [16], and Gentle and Smith originally proposed the particle-polymer metamaterial [17] as described in other patents as well [18, 19]. To further perfect the solar reflection at daytime, these radiative coolers were advanced with the silvered reflectors [20–25] and the porous scattering structures, the latter of which was also aimed towards improving on existing white cool-roof paints [26]. In general, the radiative cooler is designed to minimize the parasitic absorption while maximizing the thermal emissivity.

Although solar heating and radiative cooling are two opposite physical processes, they can be combined into the same devices for utilizing the solar energy and deep-sky simultaneously [27–29]. The radiative temperature regulator (RTR) is a representative of this kind of device, which might as well be thought of as a self-adaptive thermostat that heats itself by collecting solar energy in cold weather but cools by radiative cooling in hot weather [30]. The switching functionality is ascribed to the insulator-metal transition (IMT) mechanism by which the phase-changing materials undergo a dramatic increase in thermal emissivity at mid-infrared as the temperature reaches its critical point [31–34]. The most widely investigated IMT material is vanadium dioxide ($VO_2$), the transition temperature of which is 68°C intrinsically, and can be artificially tailored over a wide range by doping, strain-inducing, and nanostructuring [35–37]. The tunability and dynamic feature enable the radiative RTR to find various promising applications, such as building energy-saving, spacecraft thermal control, personal thermal management, and infrared camouflage.

However, realizing the passive RTR is challenging since it relies on a precise balance of solar heating and radiative cooling, unlike the dynamic radiative coolers based solely on the radiative cooling effect [32–34,38–40]. For instance, Ono et al. have invented a tandem photonic structure that can self-adaptively turn 'on' and 'off' radiative cooling based on the ambient temperature [33], which initiates a pioneering endeavor toward new functionalities of radiative cooling, as well as providing a promising solution to simultaneously utilizing the solar energy and radiative cooling. Nevertheless, the relevant work has not discussed the realization of self-heating functionality, especially the capability of switching 'cooling' and 'heating' modes for RTR. Recently, a $VO_2$-based configuration is proposed, which paves an ingenious way towards passive RTR enabled by phase-changing materials; however, due to the idealistic permittivity that has been employed for the phase-changing material, the solar heating effect is underestimated [30]. While the proposed design may work for a specific data set for $VO_2$, care has to be taken as the samples with a high absorption coefficient could lead to an inoperative photonic RTR. Therefore the critical technical point of suppressing solar parasitic absorption in the RTR has to be carefully studied. Moreover, a range of relevant issues concerning fundamental operation mechanisms and realistic system designs have to be clarified for promoting the advancement of radiative RTRs.

In this work, starting from the theoretical framework of radiative cooling, we focus especially on the core matter of RTRs, i.e., the balance of solar heating and radiative cooling, and explore the fundamental limit of the RTR with ideal selective and non-selective emissions comprehensively. Considering the relative temperature of RTRs with respect to the ambient temperature, the preference of spectral selectivity of emissivity is illustrated, and the underlying physics is explained from the thermal equilibrium perspective. The dynamic performances of the RTR with different parasitic absorption rates are also evaluated under typical weather conditions. We find that solar heating plays a crucial but subtle role in achieving the multifunctional mode of RTRs: on the one hand, if there is no solar absorption in the phase-changing device, the RTRs can switch between two modes (i.e., cooling and no-cooling modes) like [33]; on the other hand, if the solar absorption is too large, there is only one mode (i.e., heating mode) left; if and only if the solar absorption is moderate, three modes (i.e., cooling, heating, and no-cooling/heating modes) can be achieved and self-adaptively modulated with the ambient temperature. Notably, the so-called 'moderate solar absorption' is actually very small (i.e., $α_{solar}$ < 0.13 for the ideal emitter and a far smaller value for the general emitters) according to our theoretical calculations, which is far beyond the experimentally available $VO_2$ film. Besides pointing out the problem, we further give a frequency-selective photonic design based on the experimental material properties, showing that the parasitic absorption can be tailored to a moderate condition without degrading the emissivity switching for a well-performed RTR. The investigation theoretically sorts out a most important pair of competing mechanisms in RTRs, technically clears the obstacles involved in the device design, and thus provides useful guidance for implementing this kind of novel device.

**2. RESULTS AND DISCUSSION**

## 2.1 Principle of radiative RTRs

**Material requirements of RTRs.** The concept of radiative RTRs based on phase-changing material is explicitly given in Fig. 1. The phase-changing materials usually have distinct extinction coefficients ($\kappa$) at different phase-states. For instance, Fig. 1(a) gives the extinction coefficients of $VO_2$, which is a typical phase-changing material with a moderate and tunable transition point close to room temperature [41]. For instance, Wang et al. have demonstrated that the IMT temperature of as-prepared $VO_2$ can be decreased from 58°C to 5°C as the tungsten doping concentration increased from 0 wt% to 3 wt% [42]. Because the doping amount is ultra-small and mainly plays the role of inducing phase-transition, the optical property of the doped $VO_2$ is still dominated by the base material with a modified phase-transition degree. Therefore, the phase-transition temperature of $VO_2$ can be exemplified as ~17°C throughout the work for either the static or dynamic calculation. As shown in Fig. 1(a), in the insulator state, the value of $\kappa$ is relatively small across all the concerned wavelength range; however, after the insulator-metal transition, $\kappa$ changes little in the spectral range where solar irradiance accumulates but grows quite significantly in the near- to mid-infrared waveband (especially in the atmospheric window where thermal emission occurs). The dramatic change of $\kappa$ before and after phase-change indicates a strong capability of $VO_2$ on thermal modulation. On the other hand, it's noted that the absolute value of $\kappa$ in the insulator state (i.e., in the range of 0.1 to 0.5μm) is not that small, which is even comparable to that of Si widely used in solar harvesters [43]. The combination of solar absorption and thermal modulation makes $VO_2$ a most promising candidate for constructing self-adaptive temperature-control devices such as RTRs.

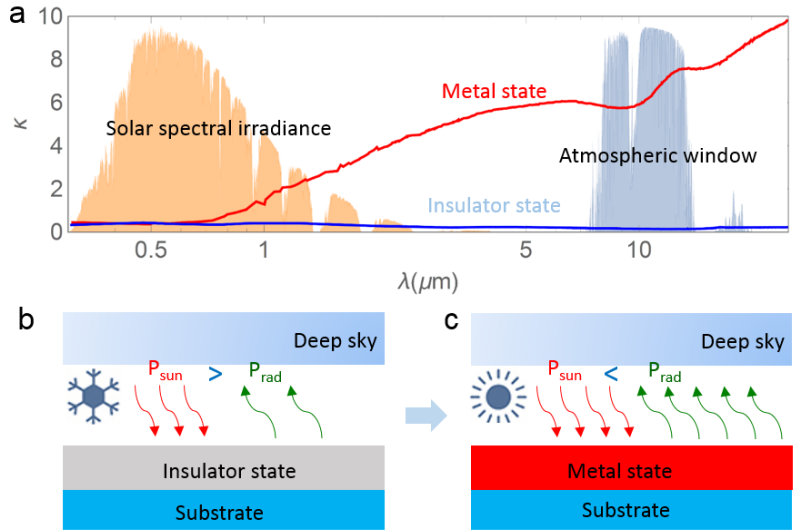

Fig. 1 Sketch for the working principle of self-adaptive thermostats based on phase-changing materials. (a) The frequency-dependent extinction coefficient ($\kappa$) of $VO_2$ in the insulator (blue curve) and metal state (red curve), indicating a strong tunability of $VO_2$ on the infrared emissivity. (b) In cold temperature, the thermostat mainly composed of $VO_2$ layer absorbs more energy from solar irradiance than it emits to deep sky (i.e., $P_{sun} > P_{rad}$) due to $VO_2$ stays at an insulator state with low emissivity, which makes the thermostat heat itself spontaneously. (c) In hot temperature, as the $VO_2$ layer changes into a metal state with high emissivity, the increment of thermal emission (green arrow) is much more than that of parasitic absorption (red arrow); therefore, the thermostat cools itself self-adaptively due to $P_{sun} < P_{rad}$.

**Working modes of self-adaptive RTRs.** Due to the phase transition is a temperature-driven process, the radiative RTRs can work in a self-adaptive way at different ambient temperatures. In cold weather, as shown in Fig. 1(b), the ambient temperature is too low to induce phase changing, and the radiated power ($P_{rad}$) of the device would be much less than the absorbed power ($P_{sun}$) from solar illumination (i.e., $P_{sun} > P_{rad}$) leading the device temperature to rise to a steady-state above the ambient temperature. In hot weather, as shown in Fig. 1(c), the phase changing occurs due to the high ambient temperature, and then $P_{sun} < P_{rad}$, which results in a temperature reduction via the so-called radiative cooling. Therefore, the temperature can be modulated in a self-adaptive manner, and the heating and cooling functions can switch as the temperature varies in different weathers or even during a day.

**2.2. Tradeoff between solar parasitic absorption and radiative cooling**

To clarify the underlying physical relationships quantitatively, we begin with the formula of net cooling power modified from the well-developed theory for radiative cooling [14],

$$P_{net}(T_{thermo}, T_{amb}, \alpha_{solar}, \varepsilon) = P_{rad}(T_{thermo}, \varepsilon) - P_{sun}(\alpha_{solar}) - P_{atm}(T_{amb}, \varepsilon) - P_{cc}(T_{thermo}, T_{amb}) \quad (1)$$

where $\alpha_{solar} = \int I_{AM1.5}(\lambda) A(\lambda) d\lambda / \int I_{AM1.5}(\lambda) d\lambda$ is the parasitic absorption ratio, $\varepsilon(\lambda)$ the spectral emissivity, $T_{thermo}$ and $T_{amb}$ the temperatures of the RTR and ambient circumstance, $A(\lambda)$ the solar absorbance, $I_{AM1.5}(\lambda)$ the standard solar irradiance, $P_{sun} = \alpha_{solar} \int I_{AM1.5}(\lambda) d\lambda \cong 900 \alpha_{solar}$ the absorbed power due to solar irradiation, $P_{rad} = \iint I_{BB}(\lambda, T_{thermo}) \varepsilon(\lambda, \theta) \cos\theta d\Omega d\lambda$ the radiated power due to thermal emission, $P_{cc} = h_c(T_{amb} - T_{thermo})$ the dissipated power due to heat conduction and convection, $h_c$ the heat transfer coefficient, and $P_{atm} = \iint I_{BB}(\lambda, T_{amb}) \varepsilon_{atm}(\lambda, \theta) \varepsilon(\lambda, \theta) \cos\theta d\Omega d\lambda$ the absorbed power due to the atmospheric opacity. The angle-dependent atmospheric emissivity ($\varepsilon_{atm}$) is related to atmospheric transmittance ($T$) by $\varepsilon_{atm}(\lambda, \theta) = 1 - T(\lambda)^{1/\cos\theta}$, where $T(\lambda)$ employs the transmittance data from the 'US standard 1976' atmosphere model with the meteorological conditions as follows: Water Column, 1762.3 atm-cm; Ozone Column, 0.34356; $CO_2$, 400 ppmv; CO, 0.15 ppmv; $CH_4$, 1.8 ppmv, Ground Temperature 300 K; Aerosol Model, Rural; Visibility, 23; Sensor Altitude, 50 km; Sensor Zenith, 180 deg [44]. It can be deduced that, by letting $T_{thermo}=T_{amb}=300K$, Eq. (1) can be simplified to the version with respect to $\alpha_{solar}$ and $\varepsilon$ since $P_{cc} = 0$, i.e.,

$$P_{net}(\alpha_{solar}, \varepsilon) = P_{rad}(\varepsilon) - P_{sun}(\alpha_{solar}) - P_{atm}(\varepsilon) \quad (2)$$

In Eq. (2), we assume that $\alpha_{solar}$ is wavelength-independent, while $\varepsilon$ is wavelength-dependent and might either be the selective or the non-selective spectrum relative to the atmospheric window. As shown in Figs. 2(a) and 2(b), the selective (non-selective) type has unit emissivity within the atmospheric window (throughout the entire wavelength range of $2.5 < \lambda < 20\mu m$) and zero emissivity elsewhere. According to Eq. (2), the contour map of $P_{net}$ versus $\alpha_{solar}$ and $\varepsilon$ is given in Fig. 2(c), which is a result corresponding to selective emissivity case and that of the non-selective case (not shown) is almost (but not exactly) the same, which is due to $P_{net}$ can degenerate at $T_{thermo}=T_{amb}$ in Eq. (2) if the the atmospheric transmittance is zero outside the atmospheric window; here the slight difference is ascribed to the non-ideal transmittance data from 'US standard 1976' model. If $T_{thermo} \neq T_{amb}$, the contour maps of two cases would become different, indicating that the two ideal spectra behave quite differently under a more frequently encountered temperature condition. As shown in Fig. 2(c), the blue and red regions respectively correspond to the dual functionalities of cooling and heating, which are bordered by the dashed line of $P_{net}=0$. In an extreme situation of $\alpha_{solar}=\varepsilon=0$, there is no absorption and emission in the system and therefore the line of $P_{net}=0$ goes through the origin; however, this is valid only for $T_{thermo}=T_{amb}$ since $P_{cc}=0$. In the situation of $\varepsilon = 0$, which corresponds to the bottom boundary of the contour map in Fig. 2(c), $P_{net} = -P_{sun}(\alpha_{solar}) < 0$; therefore, as $\alpha_{solar}$ increases, $P_{net}$ decreases, suggesting an enhanced heating effect. In the situation of $\alpha_{solar}=0$, which corresponds to the left boundary of the contour map in Fig. 2(c), $P_{net} = P_{rad}(\varepsilon) - P_{atm}(\varepsilon) > 0$; therefore, as $\varepsilon$ increases, $P_{net}$ increases and results in an enhanced cooling effect. Generally, a combination of larger $\varepsilon$ and smaller $\alpha_{solar}$ is preferred for cooling, while an inverse combination for heating. The tradeoff between $\varepsilon$ and $\alpha_{solar}$ enables the possibility of minimizing the temperature variation of RTRs under different solar irradiances.

However, as shown in Fig. 2(c), the $\alpha_{solar}$ corresponding to $P_{net}$=0 is actually fairly small, indicating the parasitic absorption of RTRs has to be carefully suppressed. To highlight the limiting factor of parasitic absorption, the relationships of ultimate temperature differences (i.e., $\Delta T=T_{thermo}-T_{amb}$) versus $\alpha$ under ideal emissivity spectra are given in Figs. 2(d) and 2(e), where $\Delta T$ is derived by letting $P_{net}(T_{thermo},T_{amb}) = 0$, $T_{amb} = 300$K and $h_c=6.9$W/(m$^2\cdot$K) as routine. As shown in Fig. 2(d), in the range of $0 < \alpha_{solar} < 0.13$, $\Delta T$ can be $> 0$ or $< 0$ as $\varepsilon$ varies from 0 to 1, which means the switch between heating and cooling is possible in such a situation. However, as $\alpha_{solar}>0.13$, the cooling mode is cut off completely, and merely the heating mode is possible, suggesting the self-adaptive RTRs function well only as $\alpha_{solar}$ is lower than the critical condition (e.g., $\alpha_{solar} \approx 0.13$ as discussed in Fig. 2(d)). The critical condition can also be observed in Fig. 2(e), which is due to that the intersection point is very close to the degenerate criteria of the selective and non-selective cases (i.e., $T_{thermo} \approx T_{amb}$) discussed in Fig. 2(c). By comparing Figs. 2(d) and (e), it can be seen that the red (blue) region in Fig. 2(d) corresponding to heating (cooling) is smaller (larger) than that in Fig. 2(e), which is apparently due to the slope of line $\varepsilon = 1$ in the former is higher than that in the latter. The underlying physics is that as $\Delta T > 0$ (i.e., $T_{thermo} > T_{amb}$), the cooling effect of non-selective emissivity is superior to that of the selective counterpart and therefore the temperature increment is retarded for the non-selective case; while $\Delta T < 0$ (i.e., $T_{thermo} < T_{amb}$), the cooling effect of non-selective emissivity is inferior to that of the selective counterpart, leading the cooling region in blue to decrease accordingly.

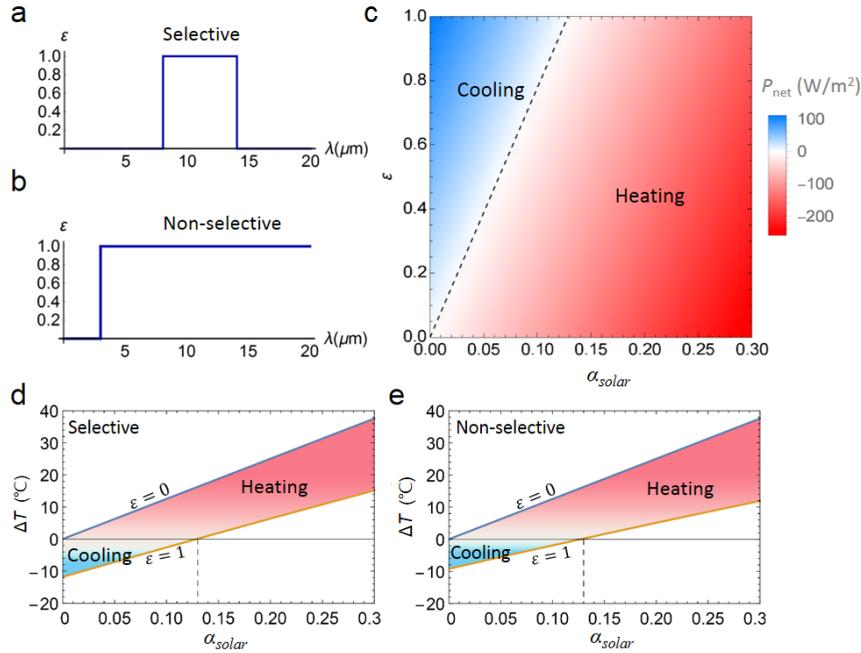

Fig. 2 Tradeoff between the parasitic absorption ratio ($\alpha_{solar}$) and the emissivity ($\varepsilon$) for achieving the functionality of cooling or heating in theoretically extreme situations. (a) The ideal selective emissivity spectrum where $\varepsilon = 1$ within the atmospheric window. (b) The ideal non-selective emissivity spectrum where $\varepsilon = 1$ across all the wavelength range beyond 2.5 μm. (c) Net cooling power ($P_{net}$) as a function of $\alpha_{solar}$ and $\varepsilon$, for both the selective and non-selective cases. The dashed lines indicate the critical condition (i.e., $P_{net} = 0$) falling in between the heating ($P_{net} < 0$)

and the cooling ($P_{net} > 0$) regimes. (d) and (e) show the influences of the parasitic absorption ratio ($\alpha_{solar}$) on the temperature difference (i.e., $\Delta T = T_{thermo} - T_{amb}$) under ultimate emissivity values ($\varepsilon = 0, 1$). The left and right columns correspond to the selective and non-selective cases, respectively.

## 2.3. Spectral preference of RTRs with moderate parasitic absorption

To determine the advantage or the application scope of the selective and non-selective emissivity spectra, the difference of net cooling powers is evaluated with an explanation from the thermal equilibrium perspective, which is crucial for exploring the ultimate capability of RTRs.

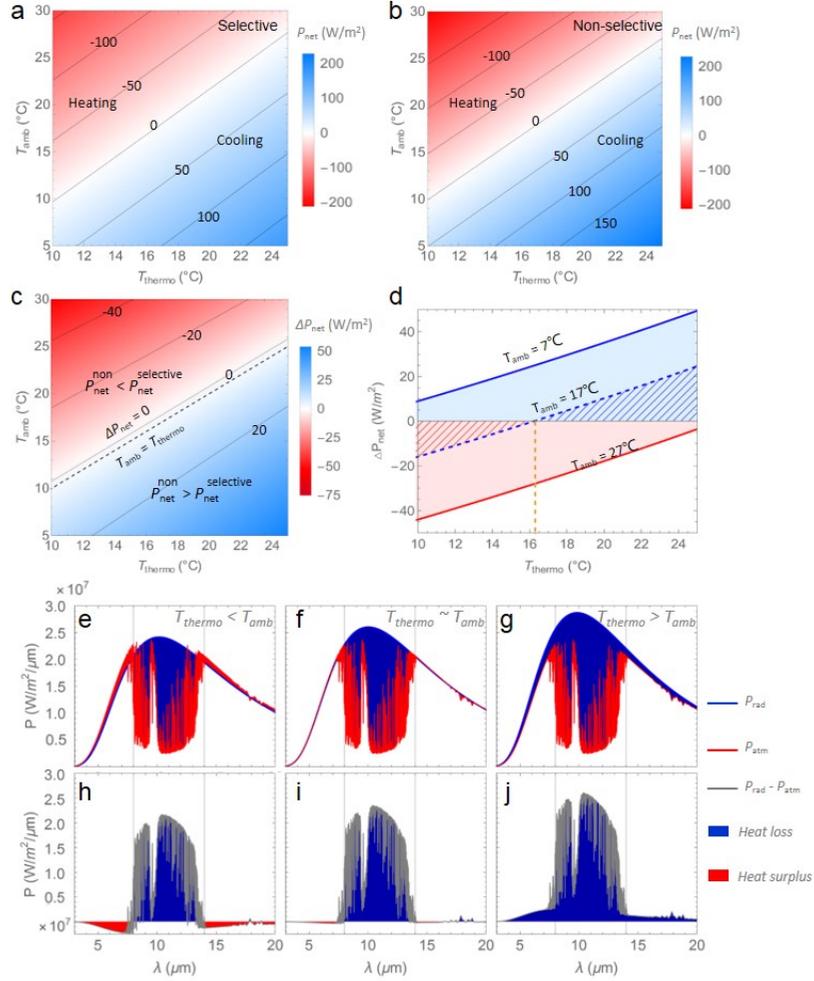

Fig. 3 Spectral preference and its underlying physics of the thermostats without considering the phase transition process. $P_{net}$ versus $T_{thermo}$ and $T_{amb}$ for (a) selective and (b) non-selective cases, respectively. (c) Contour map of $\Delta P_{net}$ versus $T_{thermo}$ and $T_{amb}$, where $\Delta P_{net} = P_{net}$ (non-selective) - $P_{net}$ (selective). (d) Line plots of $\Delta P_{net}$ as a function of $T_{thermo}$ at typical $T_{amb}$ values. (e, f, g) Spectral power densities of $P_{rad}$ (blue line) and $P_{atm}$ (red line) in the cases of $T_{thermo} < T_{amb}$, $T_{thermo} \sim T_{amb}$, and $T_{thermo} > T_{amb}$, respectively. (h, i, j) Spectral power densities of $P_{rad} - P_{atm}$ (gray line) corresponding to (e), (f), and (g), respectively. The blue and red fillings correspond to the heat loss and heat surplus, respectively. The parasitic absorption ratio is set at $\alpha_{solar} = 0.1$ in the calculations.

**Without consideration of the phase-changing effect.** Although previous works have discussed the spectral preference for general radiative cooling, they are not specifically prepared for the RTRs discussed below. Therefore, we briefly discuss it for completeness and convenience of a direct comparison with the following RTRs. According to Eq. (1), it can be concluded that once the physical parameters $\alpha_{solar}$ and $\varepsilon$ are determined, the net cooling powers (as shown in Figs. 3(a) and (b)) and their difference $\Delta P_{net}=P_{net}^{non}-P_{net}^{selective}$ (as shown in Fig. 3(c)) can be reduced to functions with respect to ambient and RTR temperatures, i.e., $\Delta P_{net}=\Delta P_{net}(T_{thermo},T_{amb})$. In the calculations, a mild weather condition (i.e., 5°C≤$T_{amb}$≤30°C) is considered, and a slightly narrower temperature range of RTRs (i.e., 10°C≤$T_{thermo}$≤25°C) is matched accordingly. Following the requirement of RTRs, a moderate parasitic absorption of $\alpha_{solar}$=0.1 is employed in the presentation. As shown in Fig. 3(c), the blue and red regions represent $\Delta P_{net}>0$ (i.e., $P_{net}^{non}>P_{net}^{selective}$) and $\Delta P_{net}<0$ (i.e., $P_{net}^{non}<P_{net}^{selective}$), respectively. The contour line of $\Delta P_{net}=0$ is located parallel to the dashed line of $T_{thermo}=T_{amb}$, the distance between which is no more than 1 °C. Fig. 3(d) selects three typical situations from Fig. 3(c) to show the interdependency between $T_{amb}$ and $T_{thermo}$ more clearly. As shown in Figs. 3(e) and 3(h), when $T_{thermo}<T_{amb}$, $P_{rad}<P_{atm}$ outside the atmospheric window, and consequently non-selective emission would bring more heat than selective emission. When $T_{thermo}≈T_{amb}$ as shown in Figs. 3(f) and 3(i), still focusing on the wavelength range outside the atmospheric window, $P_{rad}≈P_{atm}$, suggesting a balance condition for the two-type emissions. When $T_{thermo}>T_{amb}$ as shown in Figs. 3(g) and 3(j), $P_{rad}>P_{atm}$ and the non-selective emission could contribute more cooling power than the selective counterpart. In Figs. 3(e–j), with $T_{amb}$=17°C(290K), $T_{thermo}$ values are 12°C, 16.3°C, and 22°C for the scenarios of $T_{thermo}<T_{amb}$, $T_{thermo}≈T_{amb}$, and $T_{thermo}>T_{amb}$, respectively.

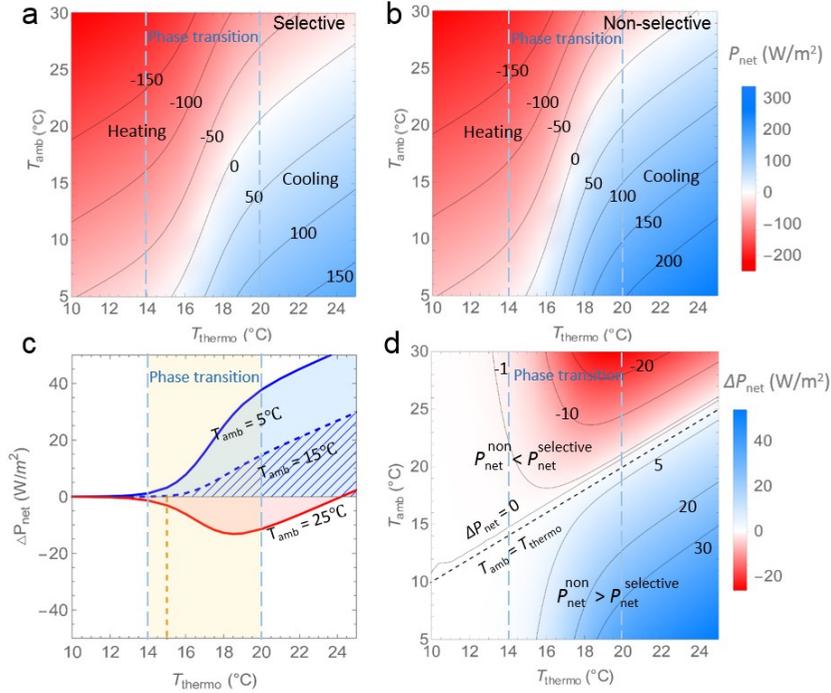

Fig. 4 Modulating capability and the spectral preference of the thermostats with considering the phase transition process. (a, b) Net cooling powers ($P_{net}$) of thermostats as a function of $T_{amb}$ and $T_{thermo}$, for the selective and non-selective cases, respectively. (c) Line plots of $\Delta P_{net}$ (i.e., $P_{net}^{non}$ - $P_{net}^{selective}$) versus $T_{thermo}$ at typical $T_{amb}$ values. (d) Contour map of $\Delta P_{net}$ as a function of $T_{thermo}$ and $T_{amb}$. A moderate parasitic absorption coefficient of $\alpha_{solar}$ = 0.1 is adopted in the calculations.

**With consideration of the phase-changing effect.** Next, we take the phase-change transition into account to investigate further the dynamic thermal modulating capability of selective and non-selective emissions. The emissivities of RTRs before and after phase-change are supposed ideally to be 0 and 1, respectively. During phase-changing, the emissivity can be treated as $\varepsilon_f=\varepsilon_{insulator}(1-f)+\varepsilon_{matal}f$ based on the effective media theory, where $f=\{1+\exp[-k(T_{thermo}-T_{pc})]\}^{-1}$ is the ratio of the metallic phase to the insulator phase, $T_{pc}=17°C$, and $k=1K^{-1}$[30]. Figs. 4(a) and 4(b) display the contour maps of $P_{net}$ versus $T_{thermo}$ and $T_{amb}$ for selective and non-selective cases, respectively. As shown in Fig. 4(a), a transition zone can be observed in the range of $14\leq T_{thermo}\leq 20°C$, which is highlighted by using a pair of dashed light-blue lines. Within the transition zone, the contour lines exhibit a non-linear correlation between $T_{thermo}$ and $T_{amb}$, indicating a retarding effect of $T_{thermo}$ due to gradually enhanced radiative cooling in the process of increasing $T_{amb}$. The contour lines are linear beyond the transition zone since the phase-changing has finished or has not yet started. For completeness, the dynamic cooling power of non-selective cases is also given in Fig. 4(b), showing a similar tendency to that of the selective counterparts. For quantitative comparisons, the differences between $P_{net}$ of selective and non-selective cases are derived at special $T_{amb}$ values. As shown in Fig. 4(c), when $T_{thermo}<14°C$, $\Delta P_{net}\approx 0$ regardless of the ambient temperature, suggesting there is almost no difference between selective and non-selective emissions. When $T_{thermo}>14°C$, as $T_{amb}$ stays much lower than the critical phase-change temperature (e.g., $T_{amb}=5°C \ll T_{pc}=17°C$), the advantage of non-selective emission is overwhelming until $T_{amb}$ approaches to $T_{pc}$ (e.g., $T_{amb}=15°C$). As $T_{amb}$ surpasses $T_{pc}$, the selective emission is preferred firstly, and then the non-selective, which becomes similar to that has been discussed in Fig. 3. The above results demonstrate that besides $T_{thermo}$, $T_{pc}$ is also a crucial factor that can be adjusted to facilitate the RTRs adapting to working conditions and emitter properties. Fig. 4(d) gives a full comparison between Figs. 4(a) and 4(b), which confirms further the conclusions in Fig. 4(c). As shown in Fig. 4(d), although the contour line of $\Delta P_{net}=0$ keeps the same as that in Fig. 3(a), the region of $\Delta P_{net}\approx 0$ (e.g., $-1<\Delta P_{net}<5W/m^2$) has significantly expanded, manifesting that the phase-changing feature highly alleviates the dependence of emission selectivity and extends the design tolerance consequently.

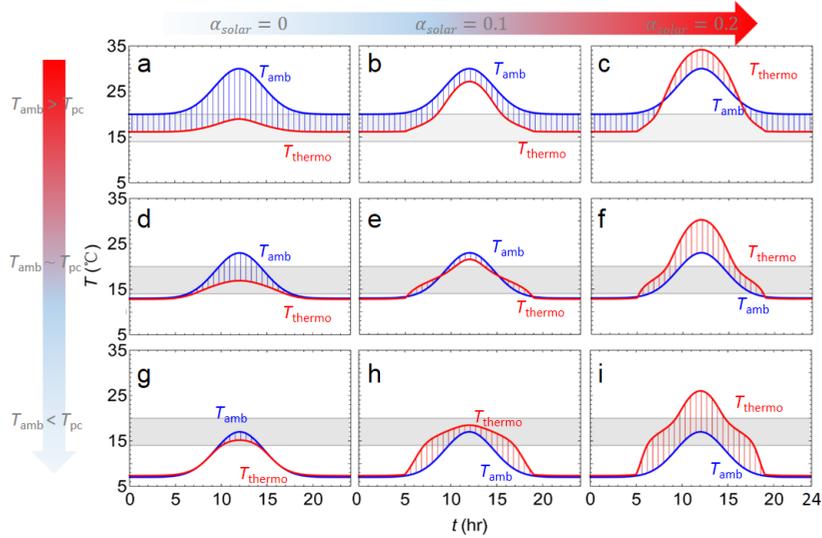

Fig. 5 The self-adaptive temperature of the VO$_2$ based thermostat ($T_{thermo}$) *with selective emissivity* as a function of the parasitic absorption ratio ($\alpha_{solar}$) in typical scenarios: (a–c) $T_{amb}>T_{pc}$; (d–f) $T_{amb}\sim T_{pc}$; (g–i) $T_{amb}<T_{pc}$. The left, middle and right columns correspond to $\alpha_{solar}=0$, $\alpha_{solar}=0.1$ and $\alpha_{solar}=0.2$, respectively. Without loss of generality, the ambient temperatures ($T_{amb}$) are modeled by the formula of $T_{amb}=T_{min}+\Delta T_{max}\exp[(t-12)^2/(-\Delta t)]$, where $T_{min}$ is the minimum

temperature of a day, $\Delta T_{max}$ is the maximal temperature difference, $\Delta t$ is the daytime period across 5:00–19:00. The temperature zone of phase-transition is set as 14 – 20°C.

## 2.4. Dynamic response of radiative RTRs with typical parasitic absorptions

The working modes of RTR during a whole day under various circumstances are further investigated. Without loss of generality, the ambient temperature is modeled by using the expression of $T_{amb} = T_{min} + \Delta T_{max}\exp[(t-12)^2/(-\Delta t)]$, where $T_{min}$ is the minimum temperature of a day, $\Delta T_{max}$ the maximal temperature difference, and $\Delta t$ the daytime period across 5:00–19:00. Fig. 5 gives the self-adaptive temperature of the VO$_2$-based RTR ($T_{thermo}$) with selective emissivity as a function of the parasitic absorption ratio ($\alpha_{solar}$) in typical scenarios: (a–c) $T_{amb} > T_{pc}$; (d–f) $T_{amb} \sim T_{pc}$; (g–i) $T_{amb} < T_{pc}$. The left, middle and right columns correspond to $\alpha_{solar} = 0$, $\alpha_{solar} = 0.1$ and $\alpha_{solar} = 0.2$, respectively. As shown in Fig. 5(a), since $T_{amb} > T_{pc}$ throughout the whole day, the RTR presents a fully metal state with ideal emissivity, and thus gains a dramatic temperature reduction of 4°C (11°C) in the nighttime (daytime). In the whole day (Fig. 5(a)), $T_{amb}$ increases largely from 20°C to 30°C, while $T_{thermo}$ varies only from 16°C to 19°C, showing a 70% reduction of temperature difference. If the weather cools down (i.e., the average of $T_{amb}$ decreases 7°C compared to that in Fig. 5(a)), as shown in Fig. 5(d), $T_{thermo}$ equals to $T_{amb}$ and is no longer reduced during nighttime due to $T_{amb}$ is beneath the phase-transition zone; however, the temperature drop in the daytime is still available but is cut down (e.g., approximately 64% at midday) in a self-adaptive manner. As the weather cools down further, it can be seen from Fig. 5(g) that the temperature reduction continues to drop, leaving only 2°C at noon. The comparison between Figs. 5(a, d, g) clearly shows that the hotter the weather is, the stronger the radiative cooling effect is (vice versa), which exactly is what the RTR needs.

Besides the ideal case of zero-heating process (i.e., $\alpha_{solar} = 0$), a moderate parasitic absorption of $\alpha_{solar} = 0.1$ is further taken into account. As shown in the middle column of Fig. 5, due to solar heating, $T_{thermo}$ in the daytime increases drastically for all the cases of $T_{amb} > T_{pc}$, $T_{amb} \sim T_{pc}$ and $T_{amb} < T_{pc}$. More importantly, a class of perfect temperature modulation mode is achieved through appropriately balancing the solar heating and radiative cooling. As shown in Figs. 5(b) and 5(h), the RTR cools down (i.e., $T_{thermo} < T_{amb}$) in hot weather (i.e., $T_{amb} > T_{pc}$) and heats up (i.e., $T_{thermo} > T_{amb}$) in cold weather (i.e., $T_{amb} < T_{pc}$), respectively. Furthermore, as shown in Fig. 5(e), the daily temperature can also be tailored promptly, i.e., the RTR cools down at noon when the temperature rises high and heats up in the morning/evening when the ambient temperature is relatively low. In addition, as shown in Fig. 5(h), the RTR heats up largely by 5°C as $T_{amb}=10$°C at 8:00 AM but by only 2°C as $T_{amb}=17$°C at noon. In brief, the self-adaptiveness of RTR based on phase-changing material can be fully exerted to modulate the temperature of different weather and even that of a day. Meanwhile, it is found that the proper parasitic absorption and phase-change temperature are also incredibly vital for realizing the stringent balance condition. If parasitic absorption breaks the limiting condition, e.g., the case of $\alpha_{solar} = 0.2$ as given in the right column of Fig. 5, the heating effect will play an overwhelming role in the regulation of temperature. More specifically, as shown in Fig. 5(c) corresponding to hot weather, $T_{thermo}$ surpasses $T_{amb}$ in the hours of 8:00–16:00, with a maximum temperature difference of 4°C above the ambient temperature at noon. As the weather becomes cooler (as shown in Figs. 5(f) and 5(i)), the increment of $T_{thermo}$ compared to the ambient becomes larger and larger as indicated by the red-line filling, showing the device has substantially turn into a self-adaptive heater.

As for the dynamic temperature of the non-selective case, its tendency is similar, but its magnitude is a little different from that of the selective case. The difference lies in the fact that while the cooling effect dominates, the temperature drop of the non-selective case is smaller than that of the selective case since $T_{thermo} < T_{amb}$; while the heating effect dominates, the temperature increment (which is inversely proportional to the cooling power) of the non-selective case is also smaller than that of the selective cases since $T_{thermo} > T_{amb}$. Therefore, it is expected that the tolerance of the non-selective emission to high parasitic absorption must be better than that of the selective counterparts. That is, if the parasitic absorption can not be well suppressed, the non-selective type might be the preferred emission spectra for RTRs.

## 2.5. Practical configurations for suppressing parasitic absorption

Finally, the possible configurations of RTRs are discussed for providing a practical solution to realizing such a self-adaptive temperature modulator. Firstly, the radiative RTR needs to possess a sharp contrast of mid-infrared emissivity before and after the phase change process for enabling a wide range of regulation on the thermal emission. To accomplish this, as shown in Fig. 6(a), a sandwich configuration of $VO_2$(20 nm)/ZnSe(1000 nm)/Ag(200 nm) is employed according to our optimization based on the experimental refractive index data from [41, 45–47], which can form an F-P cavity as the $VO_2$ layer turns into the metallic phase [48]. The physical principle is the same as the switching component of $VO_2$/$MgF_2$/W proposed by Ono et al. [33]. By using the sandwich configuration, as shown in Fig. 6(b), a high contrast emissivity is successfully achieved in the atmospheric window. However, as shown in Fig. 6(b), the parasitic absorption is too high in both the insulator ($\alpha_{solar}$ = 0.32) and metal states ($\alpha_{solar}$ = 0.44), which fails to satisfy the requirement of RTRs on parasitic absorption ratio (i.e., $\alpha_{solar}$ < 0.13). The strong parasitic absorption of the sandwich configuration is reasonable due to the F-P resonances in the visible range and the large absorption coefficient of $VO_2$ [41,49–51], which is also a well-known open technical challenge in the field of smart windows. Based on the experimental material data of $VO_2$ [37], the configuration of RTRs proposed by [26] is revisited, which proves that the structure without solar shielding could not function as an ideal RTRs efficiently, mainly due to the high parasitic absorption of $VO_2$ layers.

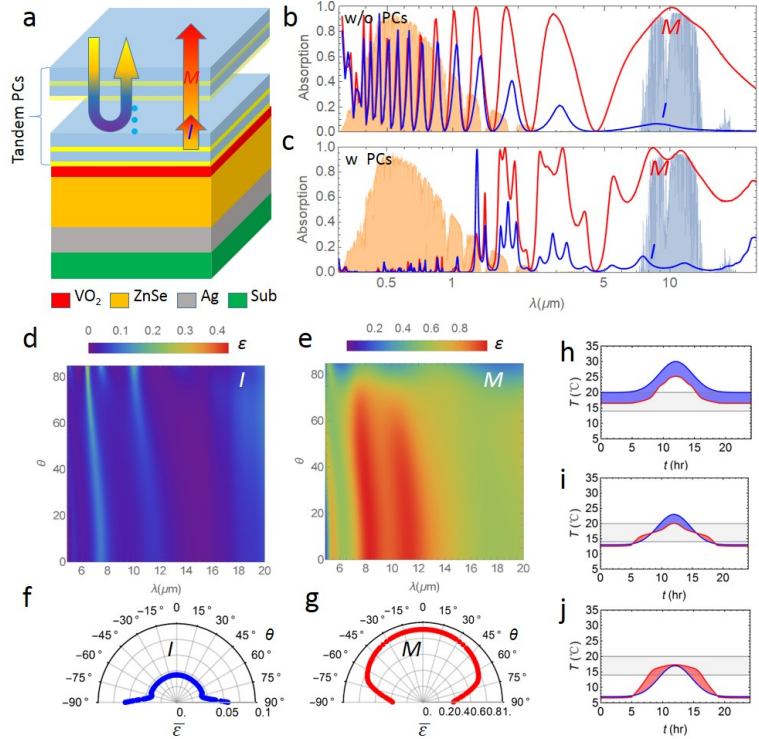

Fig. 6 An exemplary scheme for realizing $VO_2$ based thermostat with the aid of tandem photonic crystals. (a) Sketch of the conceptual configuration where the tandem photonic crystal generates extended photonic bandgap for selectively reflecting most of the solar irradiation. (b, c) Absorption/emissivity spectra across a broad waveband (from 0.3 to 20 μm), corresponding to the system without and with photonic crystals, respectively. (d, e) Azimuthal emissivity spectra of the proposed thermostat in the insulator and metal states, respectively. (f, g) Averaged

emissivity in the mid-infrared corresponding to (b) and (c), respectively. (h, i, j) The dynamic temperature of the thermostat under typical weather conditions. The modeled temperature curves and the phase transition zone are the same as those in Fig. 5. The blue and red colors indicate the cooling and heating temperature differences.

To suppress the parasitic absorption, the tandem photonic crystals (PCs) are placed on the top, which works by forming an extended photonic bandgap in the solar spectrum. The tandem PCs consist of four groups, and each group has five pairs of ZnSe and $BaF_2$ layers. We refer to also [46] for the refractive index of $BaF_2$. The optical thickness of each layer is a quarter of the center wavelength, as listed in Table 1. As shown in Figs. 6(c), with the presence of PCs, the parasitic absorption is reduced dramatically to a suitable level of $\alpha_{solar} = 0.05$ and $\alpha_{solar} = 0.06$ for the insulator and the metal states, respectively. Moreover, the emissivity is almost unchanged by the presence of PCs, since PCs are practically transparent to the mid-infrared. And the slightly increased absorption beyond 12 μm wavelengths is ascribed to the intrinsic absorption of $BaF_2$.

Besides the situation of normal incidence given in Figs. 6(b) and 6(c), the angle-dependent emissivity spectra of the device in insulator and metal states are given in Figs. 6(d) and 6(e), respectively. As shown in Fig. 6(d), the weak emissivity peaks of 8 μm and 12 μm blueshift as the angle increases, and the emissivity intensities are increased by a small amount concomitantly. The blue-shiftiness of emissivity is also observed in Fig. 6(e) for the device in metal states; however, the intensity of emissivity has not distinctly diminished until the angle reaches to 70°. The azimuthal emissivities before and after the phase change are given in Figs. 6(f) and 6(g), respectively. As shown in Fig. 6(f) for the insulator state, the average emissivity $\varepsilon_{average}$ is no more than 0.06 throughout all the angle $\theta$. As the device switches to the metal state, the average emissivity $\varepsilon_{average} > 0.95$ as $\theta = 0°$ and $\varepsilon_{average} > 0.8$ as $\theta < 60°$, which starts to drop largely until $\theta > 75°$. Due to the excellent spectral properties, it can be predicted that the proposed configuration is very suitable for RTRs. As shown in Figs. 6(h), 6(i), and 6(j), although the actual emissivity is adopted rather than the ideal emissivity as discussed before, an ideal working mode is successfully achieved, which is similar to that as discussed in Figs. 5(b), 5(e), and 5(h), respectively. As an exemplified configuration, the tandem PCs demonstrate the possibility of implementing the RTRs. To further ease the fabrication, the tandem PCs can be replaced by few-layer thin films, which can be optimized based on genetic algorithms.

Table. 1 Structure parameters of the tandem photonic crystals

| Layer groups | Materials | Center wavelength (nm) | Thickness(nm) |
| --- | --- | --- | --- |
| 1 | $(ZnSe/ BaF_2)_5$ | 400 | $(35/58)_5$ |
| 2 | $(ZnSe/ BaF_2)_5$ | 550 | $(54/90)_5$ |
| 3 | $(ZnSe/ BaF_2)_5$ | 750 | $(80/133)_5$ |
| 4 | $(ZnSe/ BaF_2)_5$ | 1100 | $(110/183)_5$ |

## 3. CONCLUSION

In summary, we have comprehensively investigated the physical process of RTRs dictated by both solar heating and radiative cooling, especially focusing on exploring the fundamental limit and practical challenge in realizing this kind of new functional device. From the theoretical formula of power conservation, the quantitative relation between the parasitic absorption rate and the average emissivity is derived, which perfectly represents the two competing aspects of parasitic absorption and thermal radiation. Unexpectedly, the parasitic absorption rate has to be small enough, even in the case with ideal emissivity, to enable the dynamic balance between heating and cooling. With a very low parasitic absorption (e.g., $\alpha_{solar} = 0$), radiative cooling always dominates the system response; with a high parasitic absorption (e.g., $\alpha_{solar} = 0.2$), solar heating dominates; only with a moderate parasitic absorption (e.g., $\alpha_{solar}$ = 0.1), the tradeoff between the two competing effects mentioned above can be controlled self-adaptively to prevent overcooling when the ambiance is cold, and thus providing a unique advantage over the general radiative coolers. However, results show that it is challenging to achieve such a moderate parasitic absorption in the $VO_2$ system. Therefore, exemplified photonic structures are designed carefully for shielding the phase-changing layer from solar illumination, which allows the realization of well-performed RTRs in this study. It is also worth noting that although the discussed RTRs can switch between the heating/cooling mode self-adaptively, its modulation magnitudes of the temperature rise and drop to the ambient are mutually restricted due to the mentioned two competing factors underlying the working principle.

**Acknowledgements**

This work was supported by the Natural Science Foundation of Jiangsu Province (BK20181167, BK20180042), National Natural Science Foundation of China (61675142, 61875143), Opening Project of State Key Laboratory of High Performance Ceramics and Superfine Microstructure (SKL201912SIC), and the Priority Academic Program Development (PAPD) of Jiangsu Higher Education Institutions.

**Disclosures**

The authors declare no conflicts of interest.